# A MACRO-ELEMENT FOR DYNAMIC SOIL-STRUCTURE INTERACTION ANALYSES OF SHALLOW FOUNDATIONS

Charisis T. CHATZIGOGOS [1], Alain PECKER[2], Jean SALENÇON[3]

## ABSTRACT

The scope of the paper is to present some aspects of the development of a "macro-element" for dynamic soil-structure interaction analyses of shallow foundations. Initially the concept of "macro-element" is introduced and is illustrated with the aid of a very simple example originating from structural engineering. Then the link is made with the modeling of the dynamic response of shallow foundations and the objectives and structure of such a tool are described with reference to the specific configuration of a circular footing resting on the surface of a heterogeneous purely cohesive soil. The principal features of the "macro-element" are then presented; the soil-structure interaction domain is reduced to a point that coincides with the center of the footing and all the (material and geometric) non-linearities are lumped at this point. A discussion on the most appropriate way to treat these non-linearities is undertaken based on experience gained with earlier works. It is suggested that the non-linearities be incorporated in the model within a unified formalism making use of the theory of multi-mechanism plasticity. Initial results concerning the definition of the ultimate surface for such a plasticity model, corresponding to the seismic bearing capacity of the foundation, are finally presented.

## INTRODUCTION

**The concept of "macro-element"**
In previous works Cremer *et al.* (2001, 2002) made use of the concept of "macro-element" as a convenient tool for a fast but concise and accurate prediction of the response of shallow foundations during time history analyses of structures. The whole issue was actually motivated by the fact that, although an extraordinary progress has taken place during recent years in terms of computational capacity and efficiency available to the engineering community, the dynamic soil-structure interaction (hereafter SSI) analyses still remain exceedingly time consuming. Among other reasons, this is also due to the fact that, in principle, non linear SSI models need to be implemented at the scale of the constituent materials of the foundation and the soil, for which appropriate constitutive laws and strength criteria are required. These models offer a detailed description of the dynamic response of the soil, the foundation and the superstructure, but they increase the size of the addressed problem considerably.

The concept of "macro-element" offers an alternative simplifying approach which reduces the size of the problem significantly while preserving the essential features of the dynamic response of the system. The concept of macro-element can be understood by introducing the following scales describing the examined soil-foundation-superstructure system:

---

[1] PhD Candidate, Laboratoire de Mécanique des Solides UMR CNRS 7649, École Polytechnique, France, Email: charisis@lms.polytechnique.fr
[2] Professor and Managing Director of "Geodynamique et Structure".
[3] Professor Emeritus.

1. The *local* scale, which is the scale of the constituent materials of the soil, the foundation and the superstructure. Elements at this scale are described by conventional constitutive laws for soil, concrete *etc*.
2. The *global* scale, which is the scale of the system in its entirety.
3. The *meso*-scale, which can be viewed as an intermediate scale between the local and the global scales. This is, for instance, the scale of some structural elements, parts of the superstructure, such as beams, columns, footings *etc*.

Within this context, the concept of "macro-element" can be viewed as a *change in scale* within the global model, where one passes from the *local* scale of the constituent materials of a specific part of the global model, say a particular structural element, to the meso-scale of this structural element viewed in its entirety. In making such a change in scale, what was originally described by a large number of elements in the local scale, now constitutes a single "macro-element" in the meso-scale, thus reducing the size of the global model significantly, which can then be treated in a much more inexpensive and efficient way.

The "macro-element", viewed simply as a part of the global model, must be described by a "constitutive law" compatible with the rest of the global model elements. This "constitutive law" must be selected in such a way so as to ensure that the response of the system, examined at the meso-scale (*i.e.* with the macro-element) correctly reproduces the features of the actual response of the model (*i.e.* at the local scale) that were retained in making the passage from the local to the meso-scale. This is an essential remark, since the passage from the local scale to the meso-scale wipes out all the characteristics of the local scale (*e.g.* stresses and displacements at any point in the soil domain near the footing *etc.*) but for those that are deemed essential for the overall behavior of the global model. The features of the system to be retained at the meso-scale model are usually defined by means of a number of generalized "stress" variables and by the corresponding generalized "strain" variables according to the type of the examined problem.

**An illustrative example**
To illustrate these ideas, a simple example from structural engineering will be presented: it concerns a steel I-beam as in Figure 1a subject to bending moment from the action of a concentrated load at midspan. The "local scale" here refers to the constituent material of the beam, *i.e.* the steel, which is assumed to be described by an elastic-perfectly plastic constitutive relationship. The solution of this problem in the "local scale" reveals the creation of a zone of plastic deformations around the central section of the beam. As the load increases, the zone of plastic deformations expands until the whole central section is plastified. On the inner and outer fibers of the beam the zone of plastic deformations has a finite width $b$. The beam cannot support any further load increase; it has reached the state of "plastic collapse".

The passage to the "meso-scale" is done by considering the "generalized" curvilinear continuous medium as in Figure 1b, which coincides with the locus of the neutral axis of the I-beam. The load increases up to its ultimate value $P_u$; the bending moment at the center of the beam equals the moment of plastic collapse $M_u$ of the central beam section and a plastic hinge is created at that point; a mechanism of plastic collapse is created and the beam can support no further load increase. It is obvious that the "plastic hinge" can be viewed as the *macro-element*, which actually represents the zone of plastic deformations in the local scale. In passing to the "meso-scale", knowledge about fibers other than the neutral axis is ignored and cannot be retrieved. Moreover, all the non-linearity is lumped at one single point, namely the plastic hinge.

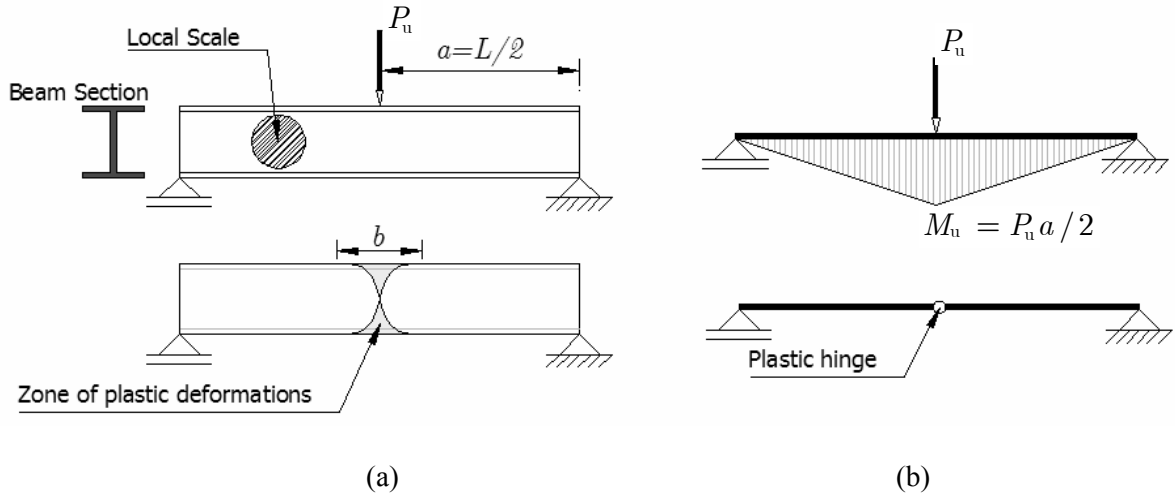

Figure 1. An elastic perfectly plastic I-beam subject to pure bending: a) Modeling at the local scale, b) Modeling at the meso-scale with a plastic hinge as a macro-element

For this system, the bending moment $M$ is the generalized stress variable (the effect of shear force is ignored), while its corresponding generalized strain variable is the curvature of the beam $\chi$. Figure 2a represents the moment – curvature diagram in the local scale; the response of the beam is linear up to a value $M_y$ which corresponds to the initiation of plastic deformations in the beam, namely on the inner and outer fibers of the beam. The beam passes to a phase of elasto-plastic response until the moment reaches its ultimate value $M_u$. Figure 2b represents the response of the system at the "meso-scale". All the non-linear part has been shrunk to one single point which corresponds to an elastic-perfectly plastic response of the beam. This fact reflects the introduction of the plastic hinge as a macro-element, which replaces the entire zone of plastic deformations in the beam.

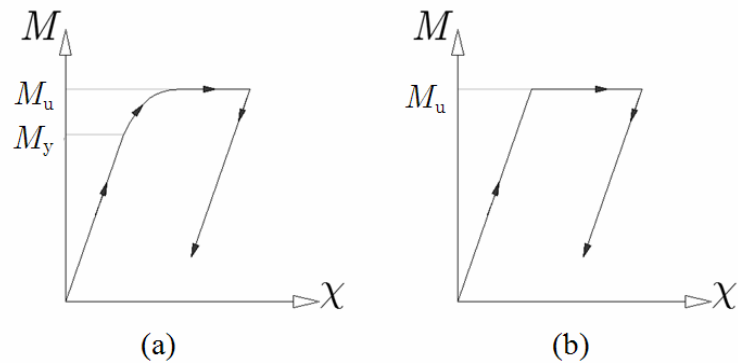

Figure 2. a) The curvature-moment diagram in the local scale and b) the idealized elastic-perfectly plastic curvature-moment diagram in the "meso-scale"

The solution of the problem of the elasto-plastic response of the beam has thus been simplified by the introduction of the generalized curvilinear continuum and the plastic hinge, as a macro-element, replacing the zone of plastic deformations in the beam. The elasto-plastic constitutive relations for the generalized curvilinear continuum can be written with respect to the generalized stress and strain variables as follows:

- $\dot{\chi} = \dot{\chi}^{el} + \dot{\chi}^{pl}$ (elastic and plastic components of total rate of curvature).

- If $|M| < M_u : \dot{\chi} = \dot{\chi}^{el} = \dfrac{K}{\dot{M}}$, with $K$ denoting the slope of the elastic branch of the curvature-moment diagram.

- If $|M| = M_u, \dot{M} < 0 : \dot{\chi} = \dot{\chi}^{el} = \dfrac{K}{\dot{M}}$

- If $|M| = M_u, \dot{M} = 0 : \begin{cases} \dot{\chi} = \dot{\chi}^{pl} = \dot{\lambda}, \ \dot{\lambda} \geq 0, \ \text{if } M = M_u \\ \dot{\chi} = \dot{\chi}^{pl} = -\dot{\lambda}, \ \dot{\lambda} \geq 0, \ \text{if } M = -M_u \end{cases}$

The above set of equations can be integrated along a prescribed loading path and all through the generalized curvilinear continuum representing the beam and allow for the determination of quantities such as the vertical deflections or the rotations of the beam. The calculation of such quantities is evidently much simpler and quicker in the examined "meso-scale" model where a "macro-element" (*i.e.* the plastic hinge) was introduced, than in any "local scale" model that would contain the geometry of the sections and would be supplied with a set of constitutive elasto-plastic equations for steel.

**The "macro-element" in the literature**
Although, the concept of "macro-element", as described above, has been extensively used in structural engineering (by the construction of "generalized media" for beams, membranes, plates, shells *etc.*), its use in geotechnical engineering is up to now rather restricted. The term "macro-element" was initially introduced by Nova & Montrasio (1991) in their study of the settlements of shallow foundations on sand; they considered the foundation and the soil as a macro-element for which the loading is expressed by a number of generalized stress variables and the displacement and rotation of the footing by the corresponding generalized strain variables. A set of incremental "plasticity-type" constitutive equations were introduced to link the generalized stress and strain variables. The initial model by Nova & Montrasio was developed for quasi-static monotonic loading. Pedretti (1998) extended the model so as to describe quasi-static loading – unloading cycles more efficiently. In parallel, Paolucci (1997) proposed a numerical tool based on the model by Nova & Montrasio permitting the study of the response of simple structures subject to dynamic (seismic) loading and taking into account the coupling between the non-linear response of the soil-foundation system and the response of the superstructure. The macro-element was further extended by Cremer *et al.* (2001, 2002) by the consideration of all the material and geometric non-linearities at the soil-footing interface (to be discussed later on), the coupling between them and their coupling with the response of the superstructure. The macro-element of Cremer *et al.* (2001, 2002) was developed for strip footings. Similar applications of the concept of "macro-element" have been developed for foundations of offshore platforms subject to quasi-static cycles of loading, as the model by Houlsby & Cassidy (2002). Nova & di Prisco (2003) presented further applications of the macro-element in problems of rock impact on the ground, soil-pipeline interaction problems *etc*. In parallel, Muir Wood & Kalasin (2004) presented a macro-element model for the dynamic response of gravity walls.

## PRESENTATION OF THE "MACRO-ELEMENT" FOR SHALLOW FOUNDATIONS

**The examined configuration**
The macro-element described in this study refers to a generic configuration presented in Figure 3. The configuration consists of a very simple superstructure of mass $m$, which is lumped at a single point. The foundation of the superstructure is composed of a circular perfectly rigid footing of diameter $D$ and mass $m_0$, which is resting on the surface of the soil. In terms of its strength criterion, the soil is considered to be purely cohesive with a cohesion increasing linearly with depth. The structure is assumed to be excited by the propagation of a seismic wave inducing the development of inertial forces in the horizontal direction and causing the excitation of the whole soil-foundation-superstructure system. The footing is thus subject to an inclined and eccentric force originating from the dynamic response of the superstructure.

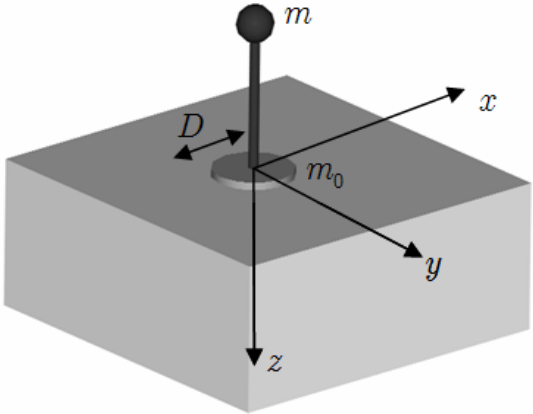

**Figure 3. Examined configuration for the development of the macro-element**

Figure 4 presents the results from a typical finite element analysis at the "local scale" with the footing being subject to an inclined and eccentric force. The results highlight the localization of plastic deformation in a zone around the footing corresponding to a bearing capacity failure with uplift of the footing. Outside this zone, the deformations remain elastic. By making the link with the example of the elasto-plastic beam, the envisaged macro-element will replace the footing and the soil in a similar way as the plastic hinge replaces the zone of plastic deformations in the beam.

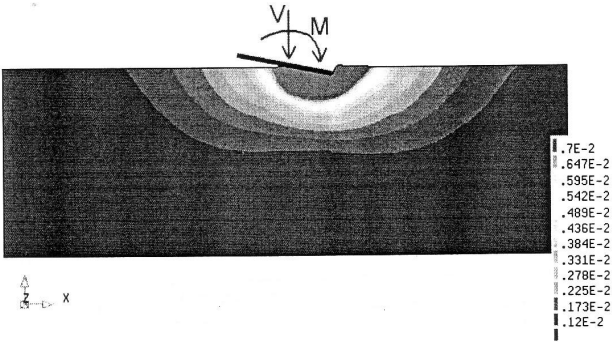

**Figure 4. Results of a footing – soil system in the "local scale". Creation of a bearing capacity failure (unconfined plastic flow) with uplift of the footing.**

**The generalized variables**
The proposed macro-element will replace the footing and the foundation soil in the global model; the entire soil domain and the footing will be reduced to a single point to which the macro-element will be attached. The selected point is the center of the footing. Since the footing has been considered perfectly rigid, its motion can be described by the motion of its center. Introduction of the macro-element implies that information at a local scale within the soil is lost and cannot be retrieved.

*Dimensionless variables*
The description of the motion of the footing center is accomplished by the introduction of a system of generalized variables. To simplify the presentation, we consider motion only in one horizontal direction (*e.g.* only within the $xz$-plane) and we introduce the resultant vertical force $N$, the horizontal force $V_x$ and the moment $M_y$ acting at the center of the footing, and the corresponding displacements: the vertical displacement $u_z$, the horizontal displacement $u_x$ and the rotation around the $y$-axis, $\theta_y$. The variables of the system are rendered dimensionless as follows:

$$\underline{F} = \begin{pmatrix} N' \\ V'_x \\ M'_y \end{pmatrix} = \frac{1}{N_{\max}} \begin{pmatrix} N \\ V_x \\ M'_y / D \end{pmatrix}, \quad \underline{u} = \begin{pmatrix} u'_z \\ u'_x \\ \theta'_y \end{pmatrix} = \frac{1}{D} \begin{pmatrix} u_z \\ u_x \\ D\theta_y \end{pmatrix} \quad (1)$$

In (1), the forces are normalized with respect to the maximum vertical force $N_{\max}$ supported by the foundation. In other words, $N_{\max}$ represents the bearing capacity of the circular footing for a centered vertical load. The generalized variables of forces and displacements are presented in Figure 5.

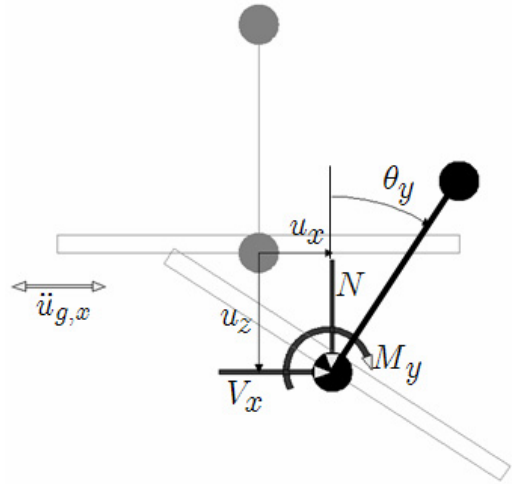

**Figure 5. Generalized forces and displacements in the macro-element**

**Scope**
Having defined the generalized forces and displacements variables that describe the macro-element, the main objective is to derive the relationship that establishes the link between them. The relationship between the generalized forces and the generalized displacements is path dependent and not invertible. However, it can be inverted if it is written in incremental form, as in the following:

$$\mathrm{d}\underline{u}_{(t)} = f\left(\underline{F}_{(t-1)}, \underline{u}_{(t-1)}, \mathrm{d}\underline{F}_{(t)}\right) \quad (2)$$

$$\mathrm{d}\underline{F}_{(t)} = f^{-1}\left(\underline{F}_{(t-1)}, \underline{u}_{(t-1)}, \mathrm{d}\underline{u}_{(t)}\right) \quad (3)$$

The above equations represent two possibilities of constructing the solution algorithm for the macro-element: (2) provides the solution for given force increments (*i.e.* at a given time step $t$, given $\underline{u}_{(t-1)}, \underline{F}_{(t-1)}, \mathrm{d}\underline{F}_{(t)}$ find $\mathrm{d}\underline{u}_{(t)}$) and (3) for given displacement increments. In the impending developments, the selection of a solution scheme with respect to the displacements (as in (3)) will be followed that will allow the implementation of the macro-element within conventional FEM codes for dynamic structural analysis.

*Loading modes*
For the types of contemplated applications, the established relationship should allow the reproduction of the foundation response under the following loading modes:

- Quasi-static monotonic loading
- Quasi-static loading - unloading cycles
- Dynamic loading (taking into account inertial and damping effects)

**Structure of the macro-element**

The structure of the envisaged macro-element will follow the scheme presented in Figure 6. The soil domain is divided in two parts: the far field, which conceptually describes the area where the response of the system remains linear, and the near field where all material and geometric non-linearities are lumped. The two fields are conceptually separated by a boundary. Accordingly, the response of the far field will be described by the linear part of the constitutive relationship in the macro-element while the response of the near field will correspond to the non-linear part of the established constitutive relationship. In the following the near and the far field properties are described.

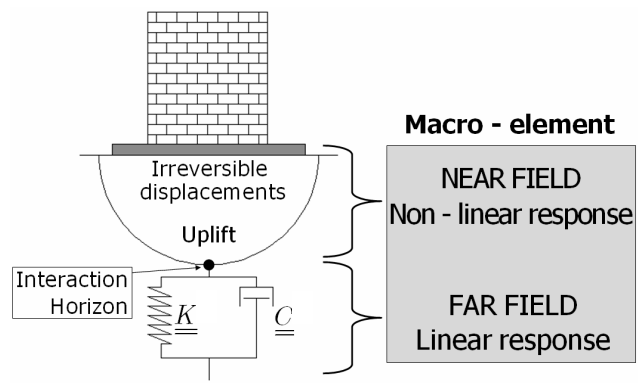

**Figure 6. Structure of the macro-element**

*Far field*

The linear response of the far field is described by a system of dashpots and elastic springs that couple all the degrees of freedom along the boundary that separates the near and the far field. For the development of the macro-element, it will be considered that the interaction horizon is reduced to a single point as in Figure 6. This point coincides with the center of the footing. The spring and dashpot attached to this point describe the "far-field" response of the system and they are supplied with the dynamic impedance (stiffness and damping) coefficients of the foundation. Since the soil response remains linear (no material damping due to soil hysteretic behavior is considered in the far field), only radiation damping will be considered. Moreover, it will be assumed that the impedances related to the considered three degrees of freedom are uncoupled. They can thus be assembled in the following stiffness and radiation damping matrices:

$$\underline{\underline{K}} = \begin{bmatrix} K_{zz} & 0 & 0 \\ 0 & K_{xx} & 0 \\ 0 & 0 & K_{\theta_y \theta_y} \end{bmatrix}, \quad \underline{\underline{C}} = \begin{bmatrix} C_{zz} & 0 & 0 \\ 0 & C_{xx} & 0 \\ 0 & 0 & C_{\theta_y \theta_y} \end{bmatrix} \qquad (4)$$

Both the stiffness and the radiation dashpot coefficients in (4) are in general frequency-dependent. Values of the coefficients in (4) for a large variety of foundation configurations and soil profiles are summarized in Mylonakis *et al.* (2006).

*Near field*

As it was mentioned, the near field response of the system is associated with all the non-linearities generated by the soil-structure interaction phenomena at the soil-footing interface. Two types of non-linearities are considered; the *material* non-linearities that refer to the development of permanent irreversible displacements of the footing due to non linear soil or interface behavior and the *geometric non-linearities* mainly referring to the phenomenon of rocking and uplift of the footing accompanied by the creation of a detachment area at the soil-footing interface.

Within the macro-element, the aforementioned sources of non-linear behavior will be addressed by means of a unified formalism based on the theory of multi-mechanism plasticity as developed by Koiter (1960) and Mandel (1965). Within the framework of this approach, one first mechanism of "plastic behavior" will describe the development of permanent displacements due to the soil non-linear behavior and a second one will describe the uplift of the footing. It is noted that when uplift occurs it is accompanied by a vertical displacement and a rotation of the footing center that may eventually be accumulated during a seismic excitation. Consequently, these two distinct mechanisms are coupled. The Koiter - Mandel theory makes it possible to treat these two coupled mechanisms simultaneously by writing them in similar forms. Using the adjective "plastic" to refer to the interaction phenomena at the soil-footing interface, the increment of plastic deformation will be expressed in both mechanisms using a plastic multiplier. It is noted that the implementation of such an approach shall constitute a significant improvement with respect to the macro-element of Cremer *et al*. (2001, 2002) where the two mechanisms where treated separately within the macro-element solution algorithm. As Cremer (2001) indicates, the adoption of the Koiter – Mandel formalism would constitute the most straightforward development for the macro-element model.

In order for the "plastic" models of uplift and of soil yielding to be implemented, three elements need to be established: a) the surface of ultimate loading or *ultimate surface,* b) the shape of the *yield surface* with the hardening laws and c) the flow rules. The yield surface for both mechanisms defines the loading combinations below which the system response is elastic whereas the ultimate surface expresses the maximum combinations of loading that can be supported by the footing, *i.e.* the *bearing capacity* of the footing. For loading combinations beyond the ultimate surface, it is assumed that an unconfined "plastic" flow occurs in the soil. Note that such a situation, although unacceptable for quasi-static loading, can be accepted for dynamic loading as long as the accumulated displacements and rotations remain below a certain limit.

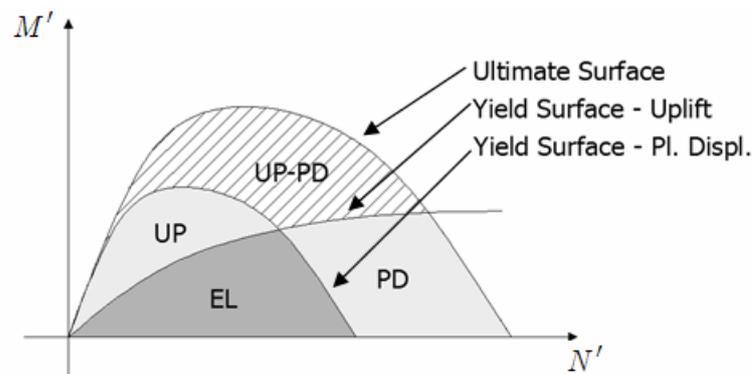

**Figure 7. Ultimate and yield surfaces for the mechanisms of uplift and the development of permanent displacements (soil yielding) with regions where each mechanism is active**

The ultimate and the yield surfaces are presented schematically in Figure 7, in the space of the loading parameters $(N', M')$, together with the region of purely elastic behavior (EL), the region where only the mechanism of uplift (UP) or only the mechanism of permanent soil displacements (soil yielding) (PD) is active and finally the region where both mechanisms are active (UP-PD). The shape and the position of the yield surfaces are changing as the loading evolves according to the hardening law. The increments of "plastic" deformation are given according to the flow rule and to whether the mechanisms are activated/deactivated at each load increment. The flow rule for both mechanisms is in general non-associative. On the contrary, the shape and position of the ultimate surface remains invariant.

*Ultimate Surface*

Up to the present stage of the development of the macro-element model, the ultimate surface has been thoroughly investigated whereas the shape and evolution of the yield surface and the flow rule will be addressed in the future. In particular, Chatzigogos *et al.* (2006) studied the maximum combinations of seismic loads of a circular footing resting on the surface of a purely cohesive soil with a vertical cohesion gradient. This was made possible using the kinematic approach of the Yield Design theory (Salençon, 2002) in which, a series of three-dimensional virtual velocity fields were examined to provide optimal upper bounds for the maximum combinations of seismic loads supported by the footing. An essential element of the study is that, besides the generalized forces $(N, V_x, M_y)$ acting on the footing, the effect of the inertial forces acting within the soil volume was taken into account through the consideration of a dimensionless parameter, function of the maximum horizontal seismic acceleration. The results have been presented in the form of surfaces in the space of the loads $(N, V_x, M_y)$ for a range of values of the inertial seismic forces in the soil or in the form of interaction diagrams between two loading parameters. Concerning the examined virtual velocity fields, each one of them corresponds to a particular virtual mechanism of bearing failure. Three large classes of virtual failure mechanisms have been considered. It is noted that the considered virtual failure mechanisms contain the case of uplift of the footing and the case of sliding along the soil-footing interface. Consequently, the ultimate surface thus established is unique for both plasticity-type mechanisms of uplift and soil yielding contained in the macro-element. A detailed description of the solution procedure, the considered virtual mechanisms of failure and the results can be found in Chatzigogos *et al.* (2005, 2006). The most interesting conclusion of the study is that the analytical expression proposed by Pecker (1997) and adopted in the Eurocode 8 for the description of the seismic bearing capacity of strip footings on homogeneous purely cohesive soils or in saturated cohesionless soils can also be used with minor modifications for circular footings in both homogeneous cohesive soils and in soils with a vertical cohesion gradient. This expression is written as follows for circular footings on homogeneous soils:

$$\frac{(1-eF_\text{h})^{c_T}(\beta V')^{c_T}}{(N')^a\left[(1-mF_\text{h}^{c_N})^{c'_N}-N'\right]^b} + \frac{(1-fF_\text{h})^{c'_M}(\gamma M')^{c_M}}{(N')^c\left[(1-mF_\text{h}^{c_N})^{c'_N}-N'\right]^d} - 1 \leq 0 \qquad (5)$$

subject to the constraints: $0 < N' \leq (1-mF_\text{h}^{c_N})^{c'_N}, |V'| \leq 1/6.06$.

In (6), we define $N', V', M'$ as in (1) and $F_\text{h} = \rho a_\text{h} D/2c$, where $D$ is the footing diameter, $c$ the soil cohesion, $\rho$ is the soil mass density, and $a_\text{h}$ a characteristic horizontal seismic acceleration. The rest of the parameters in (5) are constants and are given in Table 1.

**Table 1. Values of numerical parameters in equation (5)**

| $A=$ | 0,70 | $d=$ | 1,81 | $m=$ | 0,21 | $c_T=$ | 2,00 | $\beta$ | 2,57 |
|---|---|---|---|---|---|---|---|---|---|
| $B=$ | 1,29 | $e=$ | 0,21 | $c_N=$ | 1,22 | $c_M=$ | 2,00 | $\gamma=$ | 1,85 |
| $C=$ | 2,14 | $f=$ | 0,44 | $c'_N=$ | 1,00 | $c'_M=$ | 1,00 | | |

## CONCLUSIONS

The concept of "macro-element" stands as a convenient alternative approach for fast but concise and accurate non-linear SSI analyses for shallow foundations provided it takes the different sources of non-linearity into account. It appears that the multi-mechanism plasticity formalism would allow the simultaneous treatment of the two basic mechanisms of non-linearity (soil yielding and uplift) encountered in the problem. Initial results from the macro-element development process, which concern the definition of the ultimate surface of the "plasticity-type" model for the footing are already available from the determination of the seismic bearing capacity of the footing where it was concluded that the Eurocode 8 equation for strip footings on homogeneous soils can be extended with minor modifications to the case of circular footings on soils with a vertical cohesion gradient.

## ACKNOWLEDGEMENTS

The first author wishes to thank the École Polytechnique (Laboratoire de Mécanique des Solides) and the Public Benefit Foundation "Alexandros S. Onassis" for the financial support during the execution of this study.

## REFERENCES

Chatzigogos, C., T. "Comportement sismique des fondations superficielles : Vers la prise en compte d'un critère de performance dans la conception", Thèse de Doctorat, École Polytechnique, Palaiseau, 2007 (in preparation).

Chatzigogos, C., T., Pecker, A., Salençon, J. "Seismic bearing capacity of circular foundations", Proceedings, 1st Greece-Japan Workshop: Seismic design, observation and retrofit of foundations, Gazetas, Goto and Takashi (eds.), Athens, 141-163, 2005.

Chatzigogos, C., T., Pecker, A., Salençon, J. "Seismic bearing capacity of a circular footing on a heterogeneous cohesive soil", Soils and Foundations, submitted for publication, 2006.

Cremer, C. "Modélisation du comportement non linéaire des fondations superficielles sous séisme", Thèse de Doctorat, École Normale Supérieure de Cachan, 2001.

Cremer, C., Pecker A., Davenne L. "Cyclic macro - element of soil - structure interaction: material and geometrical non-linearities", International Journal of Numerical and Analytical Methods in Geomechanics, 25, 1257-1284, 2001.

Cremer, C., Pecker A., Davenne L. "Modelling of non linear dynamic behaviour of a shallow foundation with macro - element", Journal of Earthquake Engineering, 6, Issue 2, 175-212, 2002.

Houlsby, G., T., Cassidy, M., J. "A plasticity model for the behavior of footings on sand under combined loading", Géotechnique, 52, Issue 2, 117-129, 2002.

Koiter, W., T. "General theorems for Elastic-Plastic solids", Progress in Solid Mechanics (Sneddon, I., N., Hill, R., eds.), Vol. I, North Holland, Amsterdam, 195-221, 1960.

Mandel, J. "Généralisation de la théorie de plasticité de W. T. Koiter", International Journal of Solids and Structures, 1, 273-292, 1965.

Muir Wood, D., Kalasin, T. "Macroelement for study of dynamic response of gravity retaining walls" in "Cyclic behaviour of soils and liquefaction phenomena", Triantafyllidis, T., (ed.), Balkema Publishers, Leiden, 551-561, 2004.

Mylonakis, G., Nikolaou, S., Gazetas, G. "Footings under seismic loading: Analysis and design issues with emphasis on bridge foundations", Soil Dynamics and Earthquake Engineering, 26, Issue 9, 824-853, 2006.

Nova, R, Montrasio, L. "Settlements of shallow foundations on sand", Géotechnique, 41, Issue 2, 243-256, 1991.

Nova, R., di Prisco, C., "The macro-element concept and its application in geotechnical engineering", Fondations Superficielles, Magnan et Droniuc (ed.), Presses de l'ENPC/LCPC, Paris, 389-396, 2003.

Paolucci, R. "Simplified evaluation of earthquake induced permanent displacements of shallow foundations", Journal of Earthquake Engineering, 1, Issue 3, 563-579, 1997.

Pecker, A. "Seismic bearing capacity of shallow foundations", State-of-the-art: 11[th] World Conference on Earthquake Engineering, Acapulco, Mexico, June 23-28, 1996.

Pecker, A. "Analytical formulae for the seismic bearing capacity of shallow strip foundations", "Seismic behavior of ground and geotechnical structures", Seco & Pinto (eds.), Balkema, 261-268, 1997.

Pedretti, S. "Nonlinear seismic soil-foundation interaction: analysis and modeling method", PhD thesis, Dpt Ing Strutturale, Politecnico di Milano, 1998.

Salençon, J. "de l'Élastoplasticité au Calcul à la Rupture", Presses de l'École Polytechnique, Palaiseau, 2002.